\documentclass[aps,prd,twocolumn,10pt,reprint,nofootinbib]{revtex4-2}
\usepackage{xcolor}
\definecolor{Gray}{gray}{0.85}
\definecolor{LightCyan}{rgb}{0.88,1,1}
\usepackage[colorlinks,linkcolor={red},citecolor={blue}]{hyperref}
\usepackage{eurosym}
\usepackage{graphicx}
\usepackage{amsmath}
\usepackage{amssymb}
\usepackage{tensor}
\usepackage{amsfonts}
\usepackage{bm}
\usepackage{hhline}
\usepackage{float}
\usepackage{booktabs}
\usepackage{makecell}

\usepackage{multirow}

\def\be{\begin{equation}}
	\def\ee{\end{equation}}
\def\bea{\begin{eqnarray}}
	\def\eea{\end{eqnarray}}

\begin{document}
\title{Dark energy and a new realization of the matter Lagrangian}
\author{Shahab Shahidi}
\email{s.shahidi@du.ac.ir}
\author{Sedigheh Farahzad}
\email{farahzad@stu.du.ac.ir}
\affiliation{School of Physics, Damghan University, Damghan 41167-36716, Iran.}
\date{\today}

\begin{abstract}
	A new realization of the matter Lagrangian is introduced which models the dark energy component as a non-standard combination of thermodynamical quantities of the baryonic matter. We will prove that the present realization is independent of existing models with matter-geometry couplings and has a property that the energy-momentum tensor of both baryonic matter and dark energy is conserved separately. We further show that two possible choices of the matter Lagrangian in the $\Lambda$CDM model are not totally equivalent and investigate the background and perturbative constraints on the form of matter Lagrangian. We will also investigate cosmological implications of a test model with logarithmic DE and obtain the model parameters by confronting the model with observational data on the cosmic chronometers, Pantheon$^+$ and $f\sigma_8$ datasets. We will also explain in details the predictions of the model on the late time behavior of the universe and compare the result with $\Lambda$CDM model.
\end{abstract}
\maketitle

\section{Introduction}\label{secI}
All the observational data related to late \cite{pantheon,bao,cc} and early \cite{cmb} times favor the accelerated expansion phase in the late time universe. The first and also simplest candidate to explain this acceleration is the cosmological constant $\Lambda$, together with cold dark matter, composes the standard model of cosmology, the $\Lambda$CDM. Despite theoretical and computational simplicity of the $\Lambda$CDM model which makes it the reference model for confronting observations to theory, $\Lambda$CDM suffers from observational, phenomenological, and theoretical problems. Phenomenologically, there are the coincidence and cosmological constant problems \cite{coincidence,cosconsprob} raising the question of the true nature of the cosmological constant. Also, there are a number of theoretical issues related to the cosmological constant which are all related to the quantum corrections of the gravitational models \cite{quantuminstabilityandnaturalness}. On top of all the problems, we have a serious tension between observational data from the late time vs those extracted from the CMB observations; two of the most famous are the Hubble tension, showing about $5\sigma$ discrepancy \cite{H0tension}, and the $S_8$ tension with about $3\sigma$ \cite{s8tension} \footnote{The so-called $S_8$ tension refers to the difference of the value of the $S_8$ from weak gravitational lensing of KiDS \cite{kids} and those from the Planck data \cite{cmb}. However, the updated analysis from KiDS has reduced the discrepancy to around $0.73\sigma$, resolving the tension.}. 

All of these concerns make the $\Lambda$CDM model very difficult to be accepted as a final theory of the universe. As a result, cosmologists prefer to assume the existence of a dynamical dark energy (DE) sector responsible for the accelerated expansion of the universe. This immediately solves the phenomenological problems related to the $\Lambda$CDM and could be adjusted to reduce cosmological tensions. As a result, understanding the nature of DE is one of the fundamental challenges in modern cosmology. Dozens of generalizations of the Einstein-Hilbert action have come in recent years, willing to shed light on the true nature of DE.

Generally, one can divide modified gravities into three categories. The first and also the oldest generalization to the Einstein-Hilbert action is to make the geometry richer. This is mainly done to explain the effects of dark energy as a geometric effect. Among all, higher order models like $f(R)$ theories \cite{fr}, Weyl/Cartan theories \cite{weylcartan}, massive gravities \cite{massive}, and scalar/vector theories \cite{scalar,vector} have been extensively investigated in the literature. Also, richer geometries like Finsler spaces have become interesting recently \cite{finsler}. 
The second category deals with generalizations of the matter sector. This includes modified matter theories like $f(L_m)$, $f(T,T_{\mu\nu}T^{\mu\nu})$ and models with derivatives of matter fields \cite{modifiedmatter,derivativematter}.
The third category considers all possible couplings between the first two, i.e., the non-minimal couplings between matter and geometry; $f(R,T)$, $f(R,T,R_{\mu\nu}T^{\mu\nu})$, $f(R,T,L_m)$ etc., are basic examples \cite{nonminimalcoupling}. Phenomenologically, one can also assume the existence of DE component and then try to construct a reliable DE equation of state (eos) parameter \cite{eosparamterization}. Among all of these choices, the CPL and BA parametrizations have been very promising recently \cite{cpl,ba}. The CPL parametrization has also been generalized to contain more parameters \cite{genCPL}.

It is well-known that there are two equivalent matter Lagrangians in general relativity, namely $\mathcal{L}_m=-\rho$ and $\mathcal{L}_m=P$, where $\rho$ and $P$ are the energy density and pressure of the fluid respectively \cite{matterlagrangians}. Both Lagrangians give the ordinary matter energy-momentum tensor, which can be used to describe the matter sources in Einstein general relativity. 
%It is also known that under the energy conditions, which all the ordinary matter sources should obey, the accelerated expansion of the universe cannot be achieved, leading to the definition of dark energy. 
In this paper, we will follow the idea that the matter Lagrangian could be a combination of the two possible matter Lagrangians. In this case, the matter Lagrangian is generalized to a new function of the energy density and pressure which can account for DE. The resulting theory is independent of $f(matter)$-type theories. The reason is that the realization of matter in these theories is through the matter Lagrangian and also scalar combinations of the energy-momentum tensor. These can produce $\rho$ or $P$ and also various mixed expressions like $\rho+P$ and $\rho+3P$. But they could not provide arbitrary combinations of $\rho$ and $P$. Here, on the other hand, we assume that the matter Lagrangian could be written as
\begin{align}
	\mathcal{L}_{m,new} = f(\rho,P),
\end{align}
where $f$ is an arbitrary function. In general, the modified matter models, including the present model, has a property that the energy-momentum tensor becomes non-conserved. This non-conservation of the energy-momentum tensor can be accounted for possible matter creations from geometry and also the dark matter/dark energy interactions \cite{DMDEint}. However, the strength of this non-conservation should be very weak to become compatible with observations \cite{consonDMDEint}. As a result, the safest possibility is to modify the matter sector while keeping the conservation of the energy-momentum tensor. In this paper, we will obtain a special form of the function $f$ that ensure conservation for both ordinary and DE sectors. In the case of cosmology, we will see that the resulting function is the linear combination of the pressure and an arbitrary function of the energy density. As we have mentioned before, this form can not be obtained from $f(L_m,T)$-type models.

The paper is organized as follows. In the next section, we will compute the effective energy-momentum tensor related to the new matter Lagrangian and find a special form of the generalized Lagrangian in which the matter sector becomes conserved. We will then restrict ourselves to cosmology and find all the necessary constraints regrading the form of the function $f$, considering both background and perturbative levels. Next, we will consider the cosmological implications of a very simple example and constrain the model parameters, using the observational data from cosmic chronometers, Pantheon dataset and the $f\sigma_8$ data on the growth rate of structures. The final section will be devoted to the concluding remarks.

\section{The model}
Let us start with the action functional of the form
\begin{align}
	S = \int d^4x \sqrt{-g} \left(\frac{1}{16\pi G}\, R + \mathcal{L}_m\right),
\end{align}
where the matter Lagrangian is defined as
\begin{align}\label{matterlag}
	\mathcal{L}_m=f(\rho,P).
\end{align}
 Here $\rho$ is the energy density and $P$ is the thermodynamical pressure of the baryonic matter which we will assume to be of a perfect fluid type. It should be noted that the special cases $f=-\rho$ and $f=P$ reduces to the standard Einstein general relativity with ordinary matter sources. By defining the effective energy-momentum tensor as
\begin{align}
	T^{\text{eff}}_{\mu\nu}=-\frac{2}{\sqrt{-g}}\frac{\delta(\sqrt{-g}\mathcal{L}_m)}{\delta g^{\mu\nu}},
\end{align}
one obtains the Einstein field equation as
\begin{align}\label{eom1}
		G_{\mu \nu} = 8\pi G\, T^{\text{eff}}_{\mu \nu},
\end{align}
where $G_{\mu\nu}$ is the Einstein tensor. In order to find the energy-momentum tensor, let us first obtain the variation of the matter Lagrangian with respect to the metric tensor. We follow the process used in \cite{matterlagrangians,shahadisecondlag}. 

Let us define the particle number density $n=N/V$, the entropy per particle $s=S/N$, the enthalpy $\mu^\prime=\mu+Ts$ and the energy density $\rho=U/V$. The first law of thermodynamics and the Gibbs-Duhem relation can be rewritten as
\begin{align}
	d\rho&=Tnds+\mu^\prime dn,\\
	\rho&=\mu^\prime n-P.
\end{align}
Differentiating the Gibbs-Duhem relation results in
\begin{align}
	dP=sndT+nd\mu =nd\mu^\prime-nTds.
\end{align}
From the above relations one can see that the energy density and pressure are functions of the form $\rho=\rho(s,n)$ and $P=P(\mu^\prime,s)$. From the definition of the particle number flux and the Taub current
\begin{align}
J^\mu=\sqrt{-g}nu^\mu,\quad V_\mu=\mu^\prime u_\mu,
\end{align}
where $g$ is the metric determinant, one obtains the particle number density $n$ as
\begin{align}
n=\sqrt{\frac{g_{\mu \nu }J^{\mu }J^{\nu }}{g}},
\end{align}
where $u^\mu$ is the 4-velocity of the matter fluid with condition $u_\mu u^\mu = -1$. For conservative matter sources, the entropy density $s$, the ordinary matter number flux density vector $J^{\mu }$, and the Taub current $V_\mu$, do not depend on the metric tensor \cite{matterlagrangians}. As a result, their variations with respect to the metric identically vanish \cite{shahadisecondlag}
\begin{align}\label{assu1}
	\frac{ \delta s}{\delta g^{\alpha\beta}}=0,\quad
	\frac{\delta J^{\mu }}{\delta g^{\alpha\beta}}=0,\quad
	\frac{\delta V_\mu}{\delta g^{\alpha\beta}}=0.
\end{align}
These constraints result in the famous conservation of the the matter density flux $\nabla_\mu J^\mu=0$, and the entropy per particle flux $\nabla_\mu(sJ^\mu)=0$ \cite{matterlagrangians}.

Gathering all the above information, one can obtain the variations of the energy density and pressures as
\begin{align}\label{rhoandp}
	\delta\rho&=\frac{\rho+P}{n}\delta n,\\
	\delta P&=n\,d\mu^\prime.
\end{align}
By taking the variation of the particle number $n$ and the enthalpy $\mu^\prime$, one obtains
\begin{align}\label{14}
	\delta n&=\frac{n}{2}\left( u_{\mu }u_{\nu }+g_{\mu \nu }\right) \delta g^{\mu \nu},\nonumber\\
	\delta\mu^\prime&=-\frac12\mu^\prime u_\mu u_\nu \delta g^{\mu\nu},
\end{align}
which combining with \eqref{rhoandp} results in
\begin{align}
	\frac{\delta\rho}{\delta g^{\mu\nu}}&=\frac12(\rho+P)(g_{\mu\nu}+u_\mu u_\nu),\label{18-1}\\
	\frac{\delta P}{\delta g^{\mu\nu}}&=-\frac12(\rho+P)u_\mu u_\nu.\label{18-2}
\end{align}
Finally, one can obtain the variation of the matter Lagrangian \eqref{matterlag} as
\begin{align}
	\frac{1}{\sqrt{-g}}\delta(\sqrt{-g} \, \mathcal{L}_m) &= \frac{1}{\sqrt{-g}}\delta(\sqrt{-g} \, f(\rho,P))\nonumber\\
	&=-\frac{1}{2} \,  \, g_{\mu \nu} \, f \, \delta g^{\mu \nu} +  \, f_\rho \, \delta \rho + \, f_P \, \delta P,
\end{align}
which gives the effective energy-momentum tensor as
\begin{align}\label{emtensor1}
	T^{\text{eff}}_{\mu \nu} = (f_P - f_\rho) \, (\rho + P) \, u_\mu \, u_\nu + (f - f_\rho \, (\rho + P)) \, g_{\mu \nu}.
\end{align}
Here, subscribes denote differentiation with respect to the argument. Defining the effective energy density and pressure as
\begin{align}\label{ns}
	&P_{\text{eff}} = f - f_\rho \, (\rho + P), \\
	&\rho_{\text{eff}} = -f + f_P \, (\rho + P),
\end{align}
one can write that the effective energy-momentum tensor \eqref{emtensor1} in a perfect fluid form
\begin{align}
	T^{\text{eff}}_{\mu \nu} = (P_{\text{eff}} + \rho_{\text{eff}}) \, u_\mu \, u_\nu + P_{\text{eff}} \, g_{\mu \nu}.
\end{align}
One can easily check that the special cases $f=-\rho$ and $f=P$ gives the standard energy-momentum tensor of baryonic matter. It should be noted that since our calculations are based on the assumption that the matter source is conserved, one should have $\nabla^\nu T_{\mu\nu}=0$, where $T$ is the energy-momentum tensor of the of the baryonic matter sector. On the other hand, from the field equation \eqref{eom1}, the effective matter energy-momentum tensor $T^{\text{eff}}$ should also be conserved. This is not in general the case for \eqref{emtensor1} with arbitrary $f$. To obtain the constraint on $f$, let us define the spatial metric $h_{\mu\nu}$ as
\begin{align}
	h_{\mu\nu}=g_{\mu\nu}+u_\mu u_\nu.
\end{align}
The projection of the covariant divergence of the effective energy-momentum tensor \eqref{emtensor1} along and orthogonal to $u_\mu$ can be simplified as
\begin{align}
	(f_P-f_\rho)&(\dot\rho+3\theta (\rho+P))+(\rho+P)\dot{f}_P=0,\label{field1}\\
	(f_P-f_\rho)&(D^\alpha P+(\rho+P)\dot{u}^\alpha)-(\rho+P)D^\alpha f_\rho=0,\label{field2}
\end{align}
where we have denoted time and space derivatives as $$.\equiv u^\mu\nabla_\mu, \quad D_\mu\equiv h_\mu^{\alpha}\nabla_\alpha,$$ and defined the expansion $3\theta=\nabla_\mu u^\mu$. From these equations, it is evident that for an effective matter sector to become conserved, the following conditions must hold
\begin{align}\label{con2}
	\dot{f}_P=0, \quad D^\alpha f_\rho=0.
\end{align}
The above conditions imply that $f_P$ should be spatial and $f_\rho$ only depends on time. The consistency condition would become
\begin{align}
	\vec\nabla f_P.\frac{\partial\vec{x}}{\partial\rho}=\dot{f}_\rho\frac{\partial t}{\partial P},
\end{align}
where the time coordinate $t$ is defined along $u^\mu$ and $\vec{x}$ are the coordinates of the hypersurface described by $h_{\mu\nu}$.

Two special cases are the maximally symmetric homogeneous and isotropic space-times described by the FRW metric and the spherically symmetric static space-times. In an FLRW universe, the condition $D^\alpha f_\rho=0$ identically satisfied and the first condition in \eqref{con2} implies that
\begin{align}\label{lagcos}
	f(\rho,P)= \alpha P-B(\rho),
\end{align}
where $B$ is an arbitrary function. In order to be compatible with standard Einstein equations, we will set $\alpha=1$. The static spherically symmetric space-time on the other hand implies
\begin{align}\label{lagblack}
	f(\rho,P)= \bar\alpha \rho+\bar{B}(P),
\end{align}
where $\bar{B}$ is an arbitrary function and we set the constant $\bar\alpha=-1$ to ensure consistency with Einstein theory. It should be noted that fixing $\alpha$ and $\bar\alpha$ does not restrict the model since these constants appear as an overall factor in the theory and can be absorbed in the definition of the Newton's gravitational constant.

From the above matter Lagrangians, it is evident that the choices $\mathcal{L}_m=-\rho$ and $\mathcal{L}_m=P$ are not totally equivalent. Cosmologically, one can write the effective energy density and pressures as
\begin{align}
&P_{\text{eff}} =P-B+B_\rho(\rho+P), \\
&\rho_{\text{eff}} =\rho+B.
\end{align}
One can see that the presence of dark energy, favors the later form for the matter Lagrangian, in contrast to the standard choice in the literature. As it is evident from the calculations, the former choice would result in a non-conservative effective matter and as we have discussed, possible interactions between baryonic matter and dark energy. For these cases, one can not use the assumption \eqref{assu1}; see \cite{shahadisecondlag} for more details. It should also be noted that these Lagrangians are not equivalent in theories with non-minimal matter-geometry couplings \cite{nonequivalence}.
\section{Screening mechanism}
For a homogeneous and isotropic universe the line-element can be written as
\begin{align}\label{lineelement}
	ds^2=a^2(-dt^2+dx^2+dy^2+dz^2),
\end{align}
where $a=a(t)$ is the scale factor and $t$ is the conformal time. The universe is assuming to be filled with a perfect fluid with Lagrangian of the form \eqref{lagcos} and energy-momentum tensor as
\begin{align}\label{emt}
	T^{\text{eff}}_{\mu\nu} = (1&+B_\rho)(\rho+P)u_\mu u_\nu\nonumber\\&+\Big(P-B+B_\rho(\rho+P)\Big)g_{\mu\nu}.
\end{align}
The above energy-momentum tensor can be interpreted as a chameleon realization of dark energy through the matter Lagrangian structure itself \cite{chame}. The $B$-dependent part of the energy-momentum tensor \eqref{emt} shows an environment-dependent modification of the baryonic matter, which implements a chameleon screening mechanism. To examine this screening further, note that the effective equation of state (eos) parameter can be written as
\begin{align}\label{eosparameff}
	w_{\text{eff}} = \frac{P_{\text{eff}}}{\rho_{\text{eff}}} = -1 + (\rho+P)\frac{d}{d\rho}\ln (\rho+B).
\end{align}
For dust with $P=0$, the effective eos parameter takes the form
\begin{align}
		w_{\text{eff}} = \frac{-B+\rho B_\rho}{\rho+B}.
\end{align}
In high dense regions with $\rho\gg\rho_c$ where $\rho_c=8\pi G/3H^2$ is the critical energy density, in order for the screening to work, one should have $w_{\text{eff}}\approx0$ which restrict the function $B$ to satisfy $$B\ll \rho, \quad B_\rho \approx B/\rho.$$ For a low density regime where $\rho\sim \rho_c$ the effective eos parameter should behave like $ w_{\text{eff}}\sim -1$ which implies $$B(\rho_c) \gg \rho_c,\quad B_\rho \ll B/\rho,$$ at this limit, mimicking the cosmological constant.

It should be noted that the transition between the high and low dense regions occurs at $\rho=\rho_t$ with $B(\rho_t)\sim\rho_t$. 
\subsection{Power-law screening}
A very simple realization of the above argument can be shown to have the form
\begin{align}\label{power}
	B = \Lambda \left(\frac{\rho}{\rho_c}\right)^\alpha,
\end{align}
where $0<\alpha\ll1$. The Friedmann and Raychaudhuri equations can be obtained as
\begin{align}
	3H^2 &= 8\pi G\,a^2(\, \rho +B),\label{frid1}\\
	3H^2 + H^\prime &= -8\pi G \, a^2\big(P -B + (\rho+P)B_\rho\big).\label{frid2}
\end{align}
From the above equations, one can easily define the DE energy density and pressure as
\begin{align}\label{decomp1}
	\rho_{DE} &= B,\nonumber\\
	P_{DE} &= -B+(\rho+P)B_\rho,
\end{align}
with DE eos parameter
\begin{align}\label{eosparam}
	w_{DE} = \frac{P_{DE}}{\rho_{DE}} = -1 + (\rho+P)\frac{d}{d\rho}\ln B.
\end{align}
For the case \eqref{power}, one obtains $w_{DE}=\alpha-1$ which resembles the $\omega$CDM model of dark energy. It is well known that the value of $\alpha$ is constrained from the observational data to be $\alpha = 0.022^{+0.024}_{-0.031}$ \cite{pantheon}. This is the well-known quintessence behavior for DE.
\subsection{The logarithmic screening}
\begin{figure}[h!]
	\centering
	\includegraphics[scale=0.6]{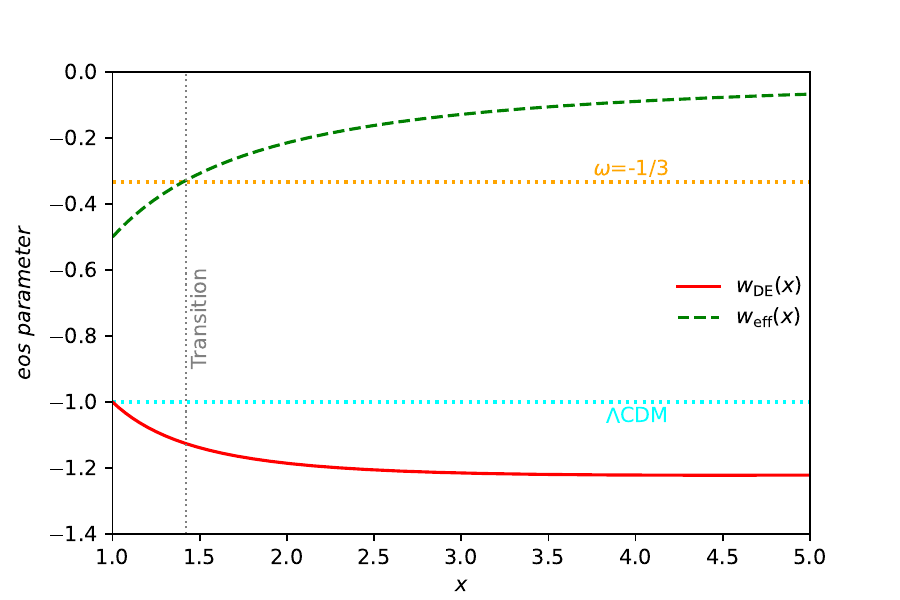}
	\caption{\label{fig1} The behavior of the effective and DE eos parameters of the logarithmic DE model. The solid line represents the DE sector and the dashed line denotes the effective part. The dotted line represents the $\Lambda$CDM behavior. Transition energy is also shown.}
\end{figure}

Another example for the realization of the above chameleon-like screening is
\begin{align}\label{logDE}
	B(\rho) = \frac{\Lambda}{2 + \ln\left(1 + \frac{\rho}{\Lambda}\right)},
\end{align}
where $\Lambda$ is an arbitrary constant. Here, the transition density is $\rho_t\sim0.42\Lambda$. One can easily prove that the above form has the property that for high dense regions with $\rho\gg\rho_t$ the effective sector behaves as dust and at low density $\rho\ll\rho_t$ reaches a constant. Here the DE eos parameter is reduced to
\begin{align}
	w_{DE} = -1-\frac{x-1}{x(2+\ln x)},
\end{align}
	where we have defined $x=1+\rho/\Lambda$. In figure \eqref{fig1} we have plotted the behavior of the effective and DE eos parameters as a function of $x$.
As one can see from the figure, the effective component tends to dust at high density $x\gg1$ and reduces to a constant at low density regime with $x\ll1$ as was discussed before. The DE component on the other hand has a cosmological constant behavior at low density regime but otherwise it behaves phantom-like. In this regard, the logarithmic DE model belongs to the phantom DE models. In figure \eqref{fig1}, we have also indicated the acceleration to deceleration phase transition line. It should be noted that the universe with logarithmic DE sector of the form \eqref{logDE} is always quintessence-like and the acceleration to deceleration phase transition occurs around the density transition domain, at $\rho=0.40\rho_t$. We will investigate the cosmological implications of this model in the following.
\subsection{The cosmological constant}
In the case of constant $B$, one can easily see that the behavior of the DE part is that of the cosmological constant with $\omega_{DE}=-1$, resulting in $\Lambda$CDM model of standard cosmology. This is though not surprising since adding a constant to the matter Lagrangian is the standard process of defining the cosmological constant.

\section{Energy conditions}
Let us now obtain possible constraints on the DE energy density $B$ such that it truly realizes the DE component. Depending on the value of $\omega_{DE}$, the DE sector could be quintessence with $-1<\omega_{DE}<-1/3$, phantom with $\omega_{DE}<-1$ or a cosmological constant $\omega_{DE}=-1$. The effective sector \eqref{ns} on the other hand, reflects the behavior of the whole universe and in line with the observational data, we will assume that the effective sector behaves like a quintessence.

First of all, the DE energy density should be positive which implies that $B>0$. In order to have a quintessence in the effective sector, the conditions
\begin{align}
	\rho_{\text{eff}}>0,\quad \rho_{\text{eff}}+P_{\text{eff}}>0,\quad \rho_{\text{eff}}+3P_{\text{eff}}<0,
\end{align}
should hold, which can be rearranged to obtain
\begin{align}\label{con1}
	-1<B_\rho<\frac{2B-(\rho+3P)}{3(\rho+P)}.
\end{align}
Noting that the baryonic matter respects all the energy conditions and using the above constraint, one can easily prove that the strong energy condition is also violated for the DE sector $\rho_{DE}+3P_{DE}<0$.

As we have mentioned before, the DE sector could behave like phantom, quintessence or can adopt a transition between the two, which usually happens in dynamical DE models \cite{eosparamterization,cpl,ba}. In the phantom regime, the necessary condition is the violation of null energy condition which can be proved to occur if we have $B_\rho<0$. The upper bound in \eqref{con1} could be negative or positive, depending on the strength of the strong energy condition for the baryonic sector. As a result one obtains the condition
\begin{align}
-1<B_\rho<\textmd{min}\left[0,\frac{2B-(\rho+3P)}{3(\rho+P)}\right],
\end{align}
for the DE sector to be phantom-like. For the DE sector to have a quintessence behavior, the function $B$ should satisfy
\begin{align}
	B>\frac{1}{2}(\rho+3P),
\end{align}
otherwise, the quintessence behavior could not be achieved. The condition \eqref{con1} then reduces to
\begin{align}
	0<B_\rho<\frac{2B-(\rho+3P)}{3(\rho+P)}.
\end{align}
In summary, the DE behavior is highly dependent on the strength of the baryonic matter and the strong energy condition $\rho+3P$. For strong enough baryons, we can not have a quintessence behavior for DE sector and the derivative $B_\rho$ should be finely constrained.
For dust with $P=0$, one can obtain the behavior of $B$ for the upper bound in \eqref{con1} as
\begin{align}
	B = B_0\rho^\frac23-\rho.
\end{align}
In this case, one can reduce the null energy conditions for the effective and DE sector to
\begin{align}
	\frac23B_0\rho^{\frac23}-\rho>0, \quad B_0\rho^{\frac23}>0,
\end{align}
which hold for sufficiently large and positive values of the constant $B_0$. The strong energy condition for the DE sector identically satisfied in this case and we obtain $\rho_\text{eff}+3P_\text{eff}=0$ which lies in the border, as was expected from the previous discussions.
\subsection{Logarithmic DE}
In the case of a dynamical DE sector of the form \eqref{logDE}, one can see that $B_\rho<0$ which implies that both null energy condition and also strong energy condition for the DE sector is violated. One can then conclude the phantom-like behavior for the DE sector. This is in line with our previous discussion and also evident from figure \eqref{fig1}.

For the effective component, one can see that for $\Lambda>0$ and $\rho_t>0$ the null energy condition is satisfied and the strong energy condition $\rho_{\text{eff}}+3P_{\text{eff}}<0$ is violated. As we have also seen in figure \eqref{fig1}, this implies a quintessence-like behavior for the logarithmic DE (LogDE) model.
\section{Matter density perturbation}
Assume a line element
\begin{align}
	ds^2 = a^2(t) \left[ -(1 + 2\Phi) \, dt^2 + (1 - 2\Psi)d\vec{x}^2 \right],
\end{align}
Here we adopt the Newtonian gauge where the scalar perturbations $E$ and $B$ vanishes and the Bardeen potentials $\Phi$ and $\Psi$ are gauge invariant.

The energy-momentum tensor of the baryonic matter can be decomposed to first order in perturbation as
\begin{align}
\delta T^0{}_0 &= -\rho\,\delta,\nonumber\\
\delta T^0{}_i &= -(1+\omega)\rho \partial_i u,\nonumber\\
\delta T^i{}_j &= c_s^2 \rho\,\delta\,\delta^i{}_j,
\end{align}
where $\omega=P/\rho$ is the eos parameter of baryons, $c_s^2=\delta P/\delta\rho$ is the sound speed, $\delta\equiv\delta\rho/\rho$ denotes the density contrast and $u$ is the scalar velocity potential.

Let us assume that the universe is filled with dust with $\omega=0=c_s^2$. Similar to the standard GR, the off-diagonal elements of the Einstein equation \eqref{eom1} gives
\begin{align}
	\Psi = \Phi.
\end{align}
Also, the $(00)$, $(i0)$ and $(ii)$ components of the linearized Einstein equation can be obtained as
\begin{align}
	&k^2\Phi+3H(\Phi^\prime+3H\Phi)=-4\pi G_{\text{eff}}a^2\rho\delta,\label{00}\\
	&k^2(\Phi^\prime+H\Phi)=-4\pi G_{\text{eff}}a^2\rho\theta,\\
	&\Phi^{\prime\prime}+3H\Phi^\prime+(2H^\prime+H^2)\Phi = 4\pi G_{\text{eff}}^\prime a^2\rho^2\delta,
\end{align}
where $\theta=-k^2 u$ is the velocity divergence and we have defined the effective gravitational constant as
\begin{align}
	G_{eff}=(1+B_\rho)G.
\end{align}
We have also Fourier transformed the spatial coordinates and $\vec{k}$ is the wave vector. 

 From the baryonic matter conservation equation $\nabla_\mu T^\mu_\nu=0$, we obtain
\begin{align}
&\delta^\prime+\theta-3\Phi =0,\\
&\theta^\prime+H\theta-k^2\Phi=0. \label{spatial}
\end{align}
By combining the above equations, one can obtain an evolution equation of the density contrast as
\begin{align}\label{del2}
		\delta^{\prime\prime} + H \delta^\prime-3(\Phi^\prime+H\Phi)+k^2\Phi= 0.
\end{align}
Now, let us work in the sub-horizon limit where $k^2\gg aH$. In this limit, one can obtain the scalar potential $\Phi$ from \eqref{00} and substituting to \eqref{del2} to obtain
\begin{align}
\delta^{\prime\prime} + H \delta^\prime-4\pi G_{\text{eff}}a^2\rho\delta =0.
\end{align}
It can be seen here that the gravitational constant in this model is matter dependent and, using condition \eqref{con1}, is always positive and smaller than the standard gravitational constant, signaling weaker effective gravity and a decrease in growth rate as we will see in the following. 

It should be noted that with the choice \eqref{lagcos}, the conservation equation of effective sector holds only in the background level. As one can see from equations \eqref{field1}-\eqref{field2}, in the perturbative level, the temporal equation still holds by applying conservation of the baryonic matter in the background and perturbative level. However, the spatial equation does not hold in the preturbative level due to the last term in equation \eqref{field2}. This part can be simplified at first order in scalar perturbation as
\begin{align}
	(\delta-3H u)\rho^2B_{\rho\rho}.
\end{align}
One can easily verify that for a slowly varying function $B$, during the matter dominated era in the sub-horizon limit, this term is negligible compared to all terms in equation \eqref{spatial}, implying that the conservation equation of the effective part holds in this limit. This is specially true for the LogDE model discussed above.

The effective sound speed can also be obtained as
\begin{align}
	c_{s,\text{eff}}^2 \equiv\frac{\delta P_{\text{eff}}}{\delta\rho_{\text{eff}}} = c_s^2+\frac{(1+\omega)\rho B_{\rho\rho}}{1+B_\rho}.
\end{align}
In order to preserve causality and avoid gradient instability, one should have
\begin{align}
	\frac{-c_{s}^2(1+B_\rho)}{(1+\omega)}<B_{\rho\rho}<\frac{(1-c_{sb}^2)(1+B_\rho)}{(1+\omega)}.
\end{align}
In matter dominated regime, one obtains $$0<B_{\rho\rho}<1+B_\rho,$$ which implies that the second derivative is positive. As a result, the effective sound speed is greater than the standard baryonic counterpart, signaling stronger pressure support.

One can also obtain the evolution of the growth rate $f=d\ln\delta/d\ln a$ as
\begin{align}\label{del1}
	\frac{df}{d\ln a}+f^2+\left(1+\frac{d\ln H}{d\ln a}\right)f=\frac32\frac{G_{\text{eff}}}{G}\Omega_m,
\end{align}
where $\Omega_m = 8\pi G a^2\rho/3H^2$ is the baryonic matter abundance. Since the function $B$ is slowly varying, the LogDE model is a small deviation of $\Lambda$CDM with $f_s = \Omega_m^{\gamma}$ and $\gamma=6/11$. By assuming 
\begin{align}
	f = f_s(1+\alpha(\Omega_m)B_\rho),
\end{align}
where $\alpha$ is an arbitrary function and substituting in equation \eqref{del1}, one can obtain the $\alpha$-equation as
\begin{align}
	2\alpha f_s^2+\left(2-\frac32\Omega_m\right)\alpha f_s = \frac32\Omega_m,
\end{align}
where we have used the matter dominated Hubble parameter 
$$\frac{d\ln H}{d\ln a}=1-\frac32\Omega_m.$$
We then obtain the effective growth rate as
\begin{align}
	f=\Omega_m^\gamma\left(1+\frac{3B_\rho}{3+2\Omega_m^{2\gamma-1}}\right).
\end{align}
The effective growth index can then be obtained as
\begin{align}
	\gamma_{\text{eff}}(a) = \gamma+\frac{3B_\rho}{(3+2\Omega_m^{2\gamma-1})\ln\Omega_m}.
\end{align}
The $\Omega_m$-dependence of $\gamma_{\text{eff}}$ implies that the relative impact of $B_\rho$ on growth increases at $\Omega_m\to 0$, even if $B_\rho$ itself remains constant. This differs from constant-coefficient parametrizations like 
$f=\Omega_m^\gamma(1+\beta B_\rho)$. It should be noted that such a behavior also appears in some $f(R,L_m)$ models \cite{2507.04079}.
It should also  be noted that in a deep matter dominated epoch the growth rate becomes $f\sim1+3/5B_\rho$ which depends on $B_\rho$ but with different factor with respect to $G_{\text{eff}}$.
\section{Cosmological implications}
Let us now consider the specific case of LogDE model presented in equation \eqref{logDE}. By defining the following dimensionless variables as
\begin{align}
	\tau &= H_0 t, \quad H = H_0 h,\nonumber\\ \bar\rho_m &= \frac{8\pi G}{3H_0^2}\rho_m,\quad
	b = \frac{8\pi G}{3H_0^2}B,\quad \Omega_d = \frac{8\pi G}{3H_0^2}\Lambda,
\end{align}
one can write the background and perturbation equations governing the Hubble parameter and the density contrast as
\begin{align}\label{back}
	h^2 = a^2 \left[\Omega_{m0}a^{-3}+\Omega_d\left(2+\ln\left(1+\frac{\Omega_{m0}}{\Omega_d}a^{-3}\right)\right)^{-2}\right],
\end{align}
and
\begin{align}\label{pert}
	\ddot\delta+h\dot\delta-\frac{1}{2a}\left(3\Omega_{m0}-b_aa^4\right)\delta = 0,
\end{align}
where $\Omega_{m0}$ is the present time baryonic matter abundance, $b_a$ is the derivative of $b$ wrt the scale factor $a$ and dot denotes derivative with respect to the dimensionless time $\tau$. We have also used the conservation equation of the baryonic matter field which as before assumed to be dust with $P=0$.

Noting that $h(a=1)=1$, one can obtain $\Omega_d$ from
\begin{align}
	2+\ln\left(1+\frac{\Omega_{m0}}{\Omega_d}\right)=\sqrt{\frac{\Omega_d}{1-\Omega_{m0}}}.
\end{align}
For integrating the matter density contrast, we will adopt the redshift coordinate defined as
\begin{align}
	1+z=\frac1a,
\end{align}
with the initial conditions taken in deep matter dominated epoch $z_*$. We will assume the same initial conditions as in $\Lambda$CDM model
\begin{align}
	\frac{d\delta}{d\ln a}\Big|_{z_*} = \delta\left|_{z_*}\right.,\qquad \delta(a_*) = a_*,
\end{align}
where $a_*$ is the scale factor at $z_*$. In this paper we will set $z_* = 7.1$.
\begin{table*}
	\centering
	\renewcommand{\arraystretch}{1.3}
	\begin{tabular}{|c||c|c||c|c|}
		\hline
		\multirow{2}{*}{\textbf{Parameter}} 
		& \multicolumn{2}{c||}{\textbf{$\Lambda$CDM}} 
		& \multicolumn{2}{c|}{\textbf{LogDE}} \\ \cline{2-5}
		& \textbf{CC+Pantheon$^+$} & \textbf{CC+Pantheon$^+$+$f\sigma_8$} 
		& \textbf{CC+Pantheon$^+$} & \textbf{CC+Pantheon$^+$+$f\sigma_8$} \\ \hline\hline
		\rule{0pt}{12pt} $\mathrm{H_{0}}$ 
		& $73.169_{-0.208}^{+0.215}$ & $73.211_{-0.195}^{+0.199}$ 
		& $73.291_{-0.202}^{+0.160}$ & $73.313_{-0.186}^{+0.194}$ \\ \hline
		\rule{0pt}{12pt} $\mathrm{\Omega_{m0}}$ 
		& $0.331_{-0.016}^{+0.017}$ & $0.324_{-0.014}^{+0.015}$ 
		& $0.366_{-0.011}^{+0.014}$ & $0.363_{-0.013}^{+0.013}$ \\ \hline
		\rule{0pt}{12pt} $\mathrm{\sigma_8}$ 
		& \text{--} & $0.744_{-0.029}^{+0.025}$ 
		& \text{--} & $0.723_{-0.027}^{+0.026}$ \\ \hline\hline
		\rule{0pt}{12pt} $\Omega_d$ 
		& \text{--} &\text{--}
		& $1.414_{-0.019}^{+0.024}$ & $1.419_{-0.017}^{+0.027}$ \\ \hline
		\rule{0pt}{12pt} $\chi^2_{red}$ 
		& $1.017_{-0.001}^{+0.001}$ & $1.013_{-0.001}^{+0.001}$ 
		& $1.020_{-0.001}^{+0.001}$ & $1.017_{-0.001}^{+0.001}$ \\ \hline
		
	\end{tabular}
	\caption{The parameter constraints of the $\Lambda$CDM and NSL models for the CC+Pantheon$^+$ and CC+Pantheon$^+$+$f\sigma_8$ dataset combinations. We have also reported the fitted value of L/Dof for completeness and also the derived parameter $\Omega_d$.}
	\label{bestfit}
\end{table*}
Let us define the observable quantity $f\sigma_8$ as
\begin{align}
	f\sigma_8(a) = f(a) \sigma(a),
\end{align}
where $f$ is the growth rate defined in the previous section and
\begin{align}
	\sigma(a) = \sigma_8\frac{\delta(a)}{\delta(1)},
\end{align}
where $\sigma_8$ is the rms amplitude of matter density fluctuations in the universe on a comoving scale of $8 Mpc$.

In order to obtain the evolution of the Hubble parameter and the growth rate, we will obtain the best fit values of $H_0$, $\Omega_{m0}$ and $\sigma_8$ by performing the Likelihood analysis using the following datasets.

\subsection{The cosmic chronometers}
The cosmic chronometers (CC) is a direct and model-independent method for determining the Hubble parameter $H(z)$ by measuring the differential age evolution of passively evolving, massive early-type galaxies. Writing the Hubble parameter as
\begin{equation}
	H(z) = -\frac{1}{1+z}\frac{dz}{dt},
\end{equation}
the value of Hubble parameter $H(z)$, can be inferred from the measurement of age difference between two nearby galaxies separated by a small redshift interval $\Delta z$. In this paper we employ the 31 CC data points \cite{cc} which assumed to be independent.
The contribution of CC dataset to the total likelihood is
\begin{align}
	\chi^{2}_{\text{CC}} = \sum_i \left( \frac{H_{\text{obs},i} - H_{\text{th},i}}{\sigma_i} \right)^2,
\end{align}
where $i$ labels the data points, $H_{\text{obs},i}$ are the observational estimates of the Hubble parameter reconstructed from differential ages, $H_{\text{th},i}$ are the theoretical predictions of the model at corresponding redshifts, and $\sigma_i$ denotes the reported 1$\sigma$ uncertainties.

\subsection{The Pantheon$^+$}
The Pantheon$^+$ compilation \cite{pantheon} represents the most updated and homogeneous collection of Type Ia supernova (SN Ia) distance measurements, consists of about 1500 spectroscopically confirmed SNe~Ia spanning the redshift range $0.001 < z < 2.26$, combining observations from 18 different surveys. Pantheon$^+$ improves the Pantheon dataset by enhancing the photometric calibration and refining light-curve fitting.
Here we employ the Pantheon$^+$ dataset with SH0ES Cepheid calibration \cite{SH0ES}. The Pantheon$^+$ measurements are not independent and the covariance matrix is provided in \cite{pantheon}. The contribution of the Pantheon$^+$ dataset to the total likelihood is 
\begin{equation}
	\chi^2_{\mathrm{Pantheon}^+}
	=
	\left[ \vec{\mu}_{\mathrm{obs}} - \vec{\mu}_{\mathrm{th}} \right]^{T}
	C^{-1}
	\left[ \vec{\mu}_{\mathrm{obs}} - \vec{\mu}_{\mathrm{th}} \right],
\end{equation}
where $C$ is the covariance matrix of the Pantheon$^+$ data.
\linebreak
\subsection{The $f\sigma_8$ data}
We utilize a compilation of $f\sigma_8$ measurements spanning the redshift range $0.02 < z < 1.95$, derived from various galaxy redshift surveys including SDSS, BOSS, eBOSS and 6dFGS \cite{fsigma8data}. The data is assumed to be independent, implying a diagonal covariance matrix.
​

The contribution of the growth rate dataset to the total likelihood is
\begin{align}
	\chi^{2}_{f\sigma_{8}} = \sum_i \left( \frac{(f\sigma_8)_{\text{obs},i} - (f\sigma_8)_{\text{th},i}}{\sigma_i} \right)^2,
\end{align}
where $i$ labels the data points.

In this paper, we will use two different combinations of the datasets, namely
\begin{itemize}
	\item CC+Pantheon$^+$,
	\item CC+Pantheon$^+$+$f\sigma_8$.
\end{itemize}
The likelihood function can then be defined as
\begin{align}
	L=L_0e^{-\chi^2/2},
\end{align}
where $L_0$ is the normalization constant with corresponding loss function for each case. 
\begin{figure*}
	\includegraphics[scale=0.3]{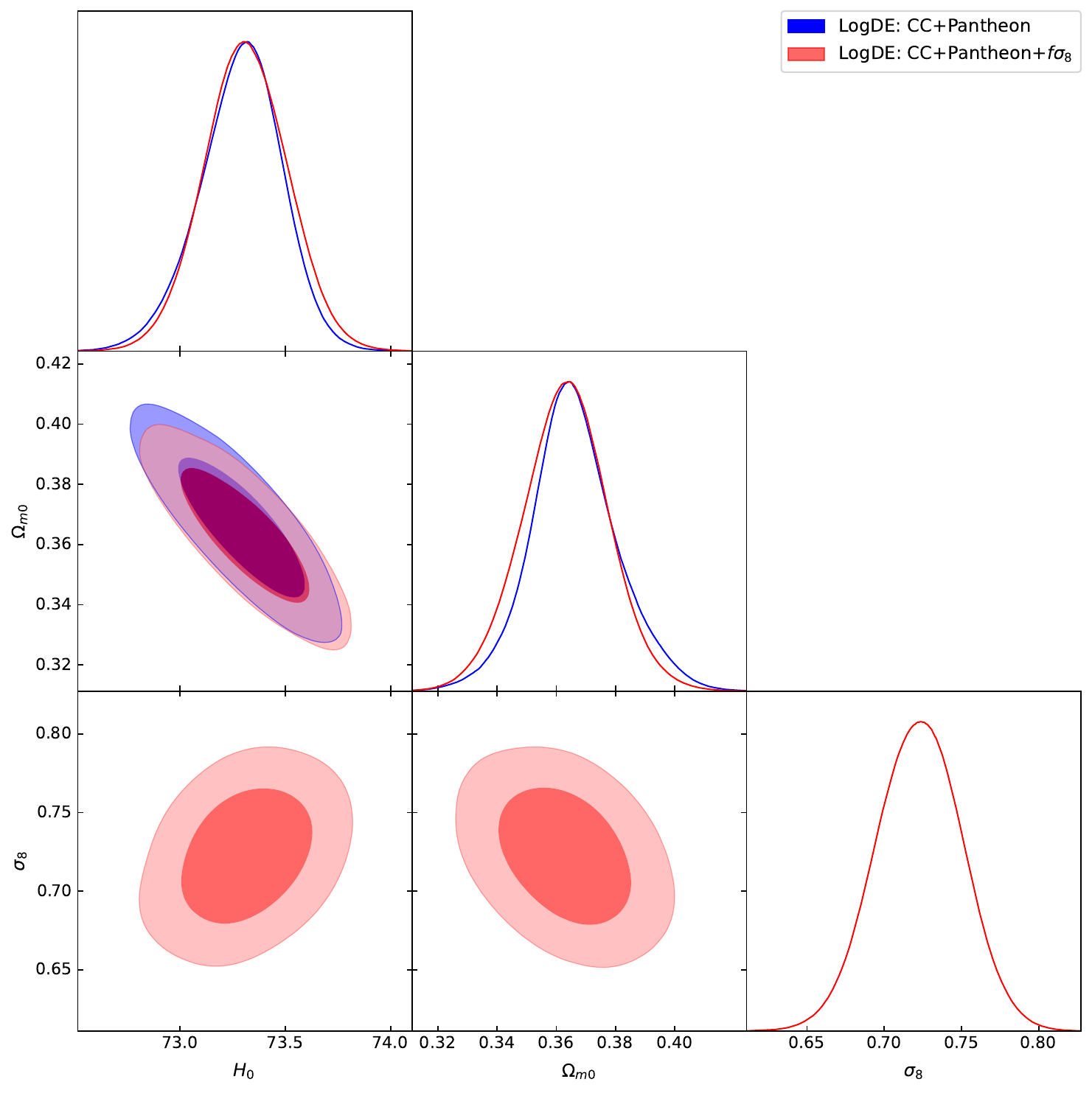}\includegraphics[scale=0.376]{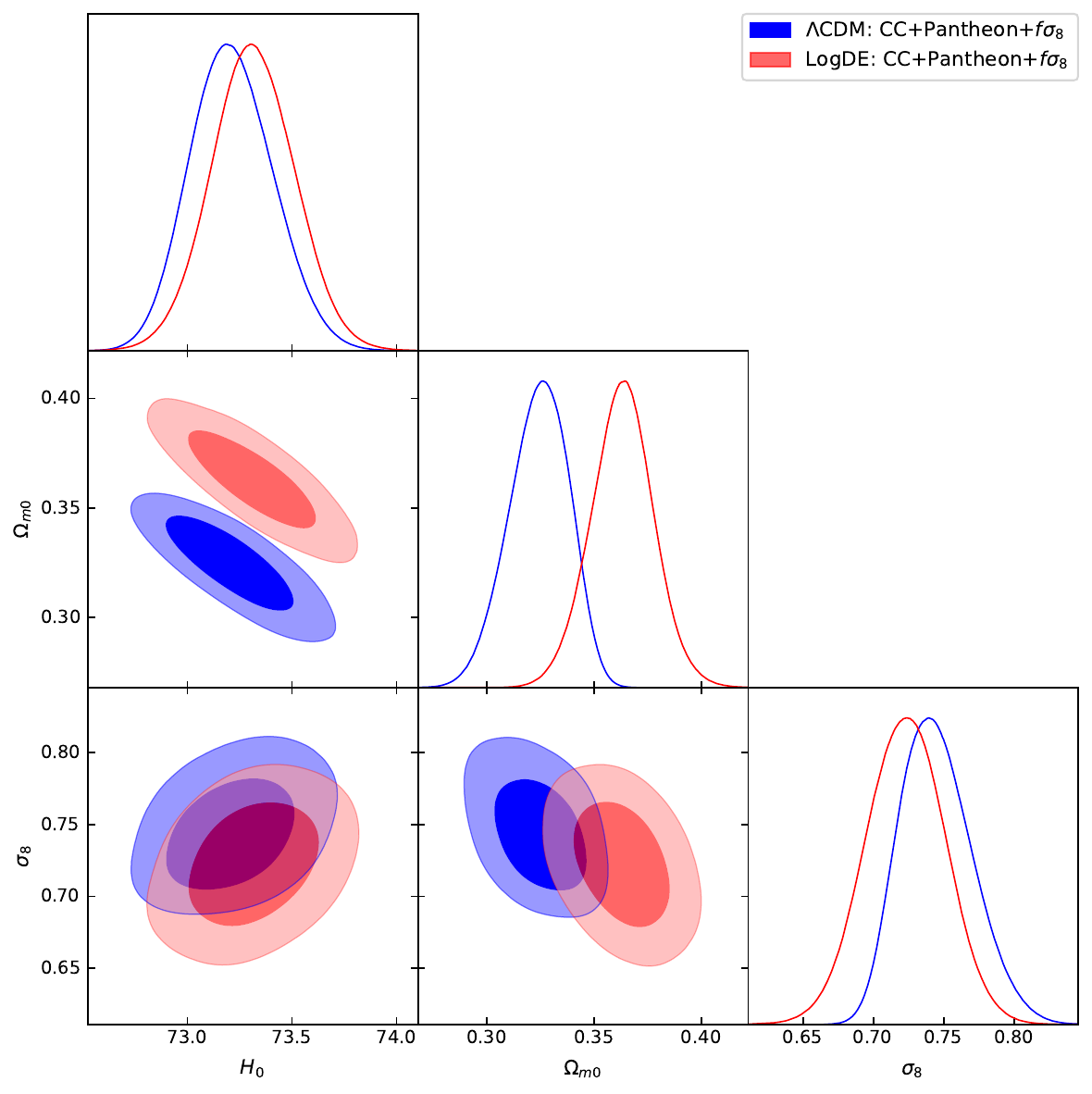}
	\caption{\label{cornerMG} The corner plot of the values of $H_0$, $\Omega_{m0}$ and $\sigma_{8}$ parameters with their $1\sigma$ and $2\sigma$ confidence levels for the logarithmic DE model (left) and together with $\Lambda$CDM model (right).}
\end{figure*}
The results are summarized in table \eqref{bestfit}. We have also included the best fit values of the parameters for $\Lambda$CDM model. Also, the constraints on the derived parameter $\Omega_d$ is shown in the table. It can be seen from the  reduced chi-squared for both datasets that the fitting process is statistically plausible.

The corner plot for the values of parameters $H_0$, $\Omega_{m0}$ and $\sigma_{8}$ with their $1\sigma$ and $2\sigma$ confidence levels for the LogDE model is shown in figure \eqref{cornerMG}. We have also reported the corner plot of the cosmological parameters for both LogDE and $\Lambda$CDM models.
It can be seen from these diagrams that $H_0$ and $\Omega_{m0}$ are correlated, similar to the $\Lambda$CDM model. However, the magnitude of the correlation between the parameters are smaller in the LogDE model compared to the $\Lambda$CDM model.
	\begin{table}[H]
		\centering
		\setlength{\tabcolsep}{11pt}      % Increases horizontal spacing
		\begin{tabular}{|c||c|c|c|}
			\hline
			\textbf{Parameter} & $\boldsymbol{H_0}$ & $\boldsymbol{\Omega_{m0}}$ & $\boldsymbol{\sigma_8}$ \\
\hline\hline
			$\boldsymbol{H_0}$ 
			& \makecell{1} 
			& \makecell{-0.711 \\ \textbf{-0.712}} 
			& \makecell{0.147 \\ \textbf{0.311}} \\
			\hline
			$\boldsymbol{\Omega_{m0}}$
			& \makecell{-0.711 \\ \textbf{-0.712}} 
			& \makecell{1} 
			& \makecell{-0.211 \\ \textbf{-0.435}} \\
			\hline
			$\boldsymbol{\sigma_8}$
			& \makecell{0.147 \\ \textbf{0.311}} 
			& \makecell{-0.211 \\ \textbf{-0.435}} 
			& \makecell{1} \\
			\hline
		\end{tabular}
			\caption{The Pearson correlation matrix for the $H_0$, $\Omega_{m0}$ and $\sigma_{8}$ parameters. In each cell, the top value corresponds to the LogDE model and the bottom value to the $\Lambda$CDM model. The value with the larger correlation is shown in bold.}
		\label{pearson}
	\end{table}
In table \eqref{pearson} we have reported the Pearson correlation matrix for both models for the dataset combination CC+Pantheon$^+$+$f\sigma_8$. It can be seen from the table that in the LogDE model the parameters are more independent compared to the $\Lambda$CDM model.

\begin{figure}[H]
	\includegraphics[scale=0.48]{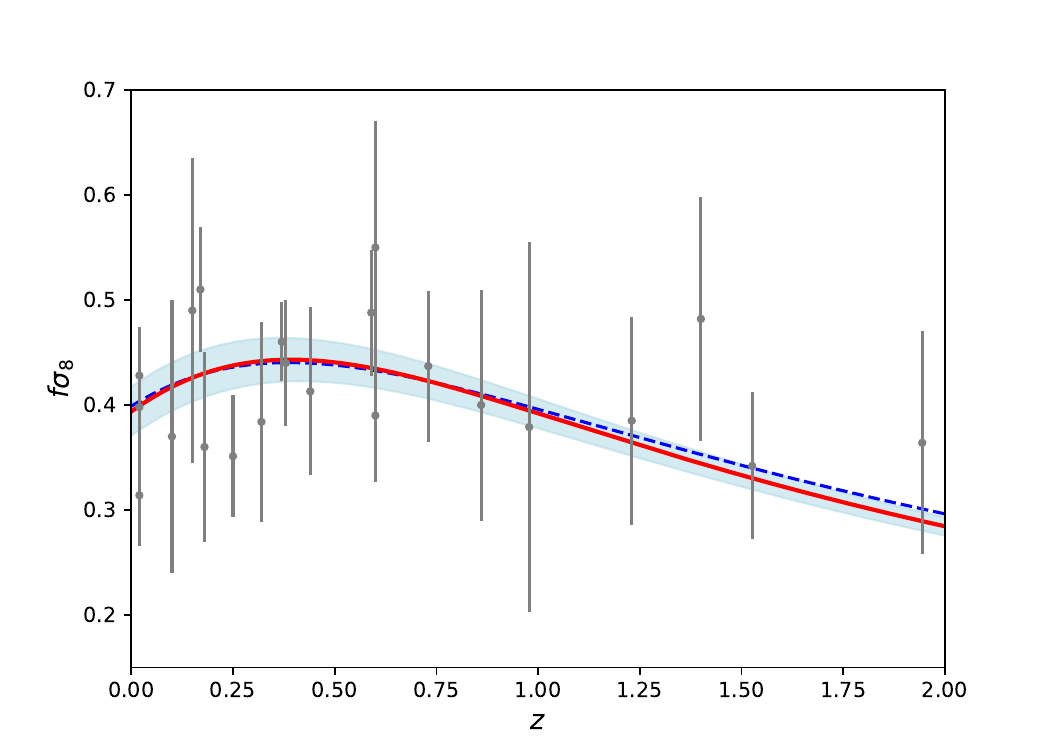}
	\caption{\label{figfsigma8} The behavior of $f\sigma_8$ as a function of $z$ for the LogDE model for the best fit values of the parameters as given by table \eqref{bestfit}. The shaded area denotes the $1\sigma$ error. Dashed lines represent $\Lambda$CDM model. The error bars correspond to the observational data.}
\end{figure}

In figure \eqref{figfsigma8} we have plotted the behavior of the $f\sigma_8$ as a function of redshift $z$ for both LogDE and $\Lambda$CDM models. 
It can be seen from the figure that the behavior of the models are very similar. The difference showed up at earlier times in which the LogDE model predicts smaller values, indicating suppressed growth of structure relative to $\Lambda$CDM model at earlier times.

\begin{figure}[H]
	\includegraphics[scale=0.48]{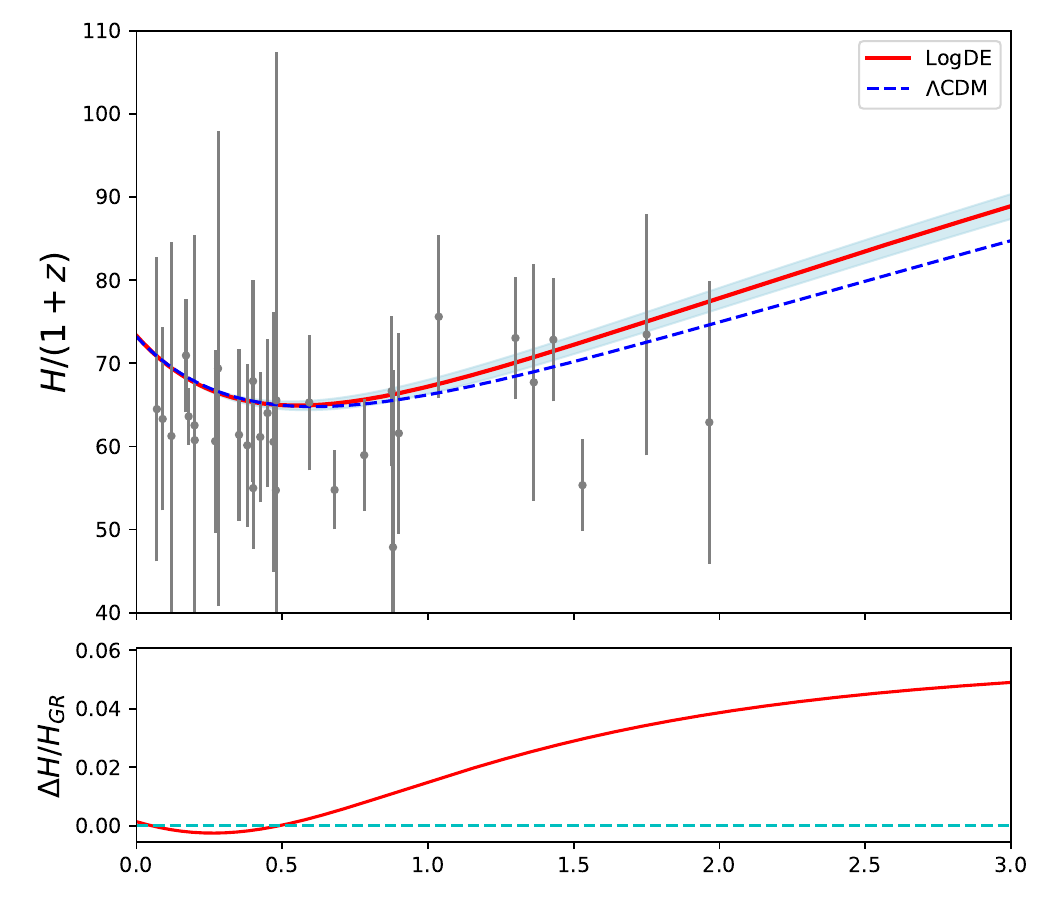}
	\caption{\label{fighubble} The behavior of the rescaled Hubble parameter $H/(1+z)$ for the LogDE model (top panel) and the difference between the LogDE and $\Lambda$CDM models (bottom panel) as a function of the redshift for the best fit values of the parameters as given by table \eqref{bestfit}. The shaded area denotes the $1\sigma$ error. Dashed lines represent $\Lambda$CDM model. The error bars correspond to the observational data of the cosmic chronometers dataset.}
\end{figure}

In figure \eqref{fighubble} we have plotted the evolution of the Hubble parameter as a function of redshift $z$. 
It can be seen from the figure that the evolution of the Hubble function is identical to the $\Lambda$CDM model at late times. The departure begins at redshifts $z\approx0.5$ where the LogDE model predicts higher values of the Hubble function compared to the $\Lambda$CDM model. This implies more rapid expansion at early times. The minimum of the reduced Hubble function represents the time of deceleration to acceleration phase transition. From the figure, one can see that the transition occurs almost at the same time as in $\Lambda$CDM model.
\begin{figure}[H]
	\includegraphics[scale=0.47]{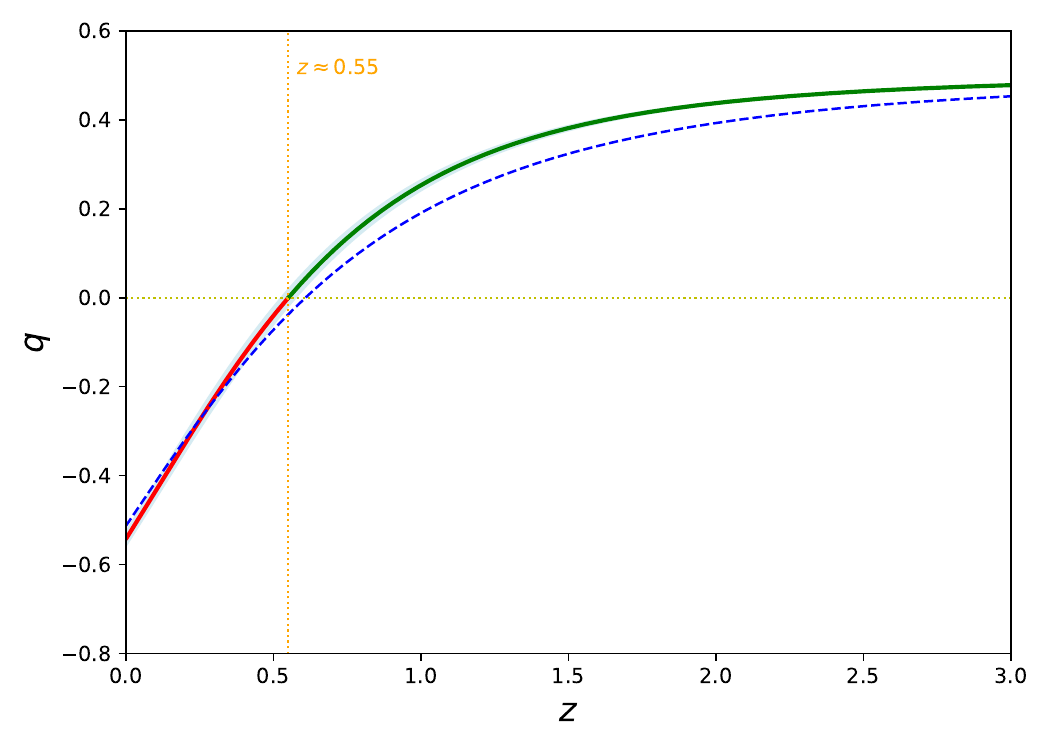}
	\caption{\label{figdec} The behavior of the deceleration parameter $q$ as a function of  redshift $z$ for LogDE model for the best fit values of the parameters as given by table \eqref{bestfit}. The shaded area denotes the $1\sigma$ error. Dashed lines represent $\Lambda$CDM model. The vertical dotted line denotes the deceleration to acceleration phase transition redshift.}
\end{figure}
In figure \eqref{figdec}, we have plotted the evolution of the deceleration parameter as a function of redshift $z$. One can see that the LogDE model predicts a slightly larger acceleration at the late times and also larger deceleration at earlier times. The deceleration to acceleration phase transition takes place at $z\approx0.55$ which is very close to the $\Lambda$CDM value $z_{\Lambda}=0.609$. This implies that the acceleration epoch in the LogDE model is a little younger than its $\Lambda$CDM counterpart. This is in fact shows that the LogDE is slightly stronger than the cosmological constant.
\begin{figure}[H]
	\includegraphics[scale=0.5]{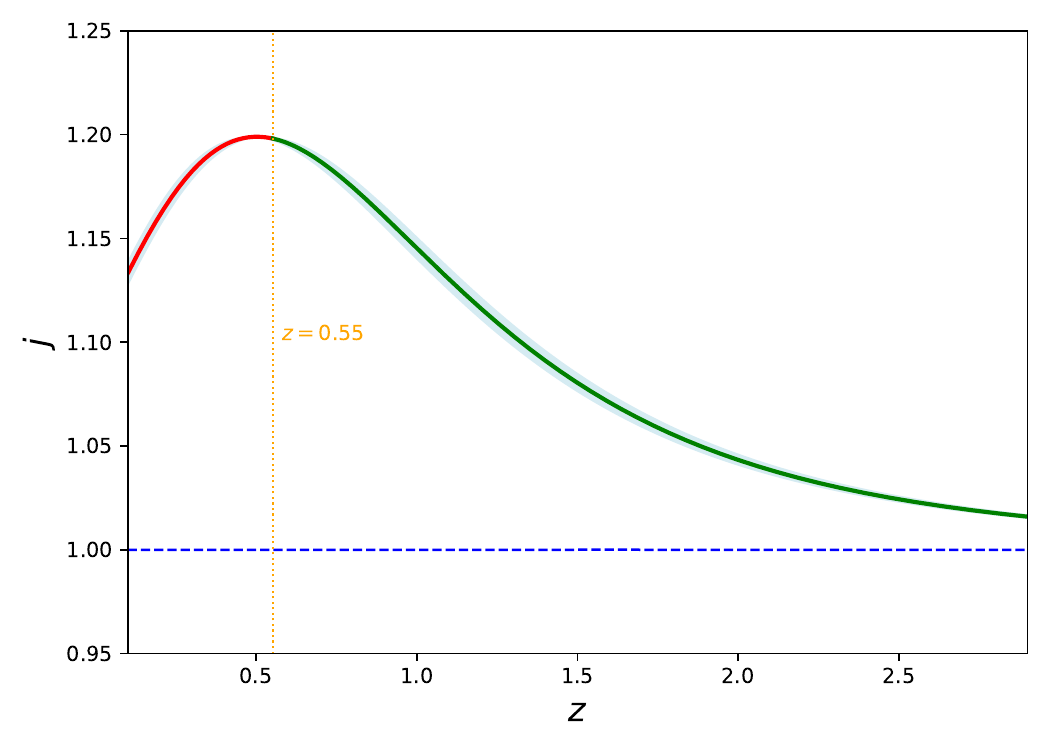}
	\caption{\label{figjerk} The behavior of jerk as a function of redshift $z$ for the LogDE model for the best fit values of the parameters as given by table \eqref{bestfit}. The shaded area denotes the $1\sigma$ error. Dashed lines represent $\Lambda$CDM model.}
\end{figure}
As one can see from the Hubble diagram \eqref{fighubble}, the concavity in the LogDE model is greater than the $\Lambda$CDM model. This can be seen quantitatively from the jerk parameter, defined as a third derivative of the scalar factor and can be expressed in terms of the deceleration parameter as
\begin{align}
	j(z) = 2q^2 + q + (1 + z)\frac{dq}{dz}.
\end{align}
In figure \eqref{figjerk} we have plotted the evolution of the jerk parameter $j$ as a function of the redshift $z$. In $\Lambda$CDM model, one can easily prove that the jerk parameter is always equal to one, so any deviations from the unity is the characteristic of DE model. For the LogDE model, one can see that the value of the jerk parameter is always greater than unity signaling greater concavity on the rescaled Hubble parameter as we have discussed before. It should also be noted that the value of the jerk parameter tends to the $\Lambda$CDM value at earlier times, showing the weakening of the DE sector. This is in line with our previous arguments on the model.

\begin{figure}[H]
	\includegraphics[scale=0.5]{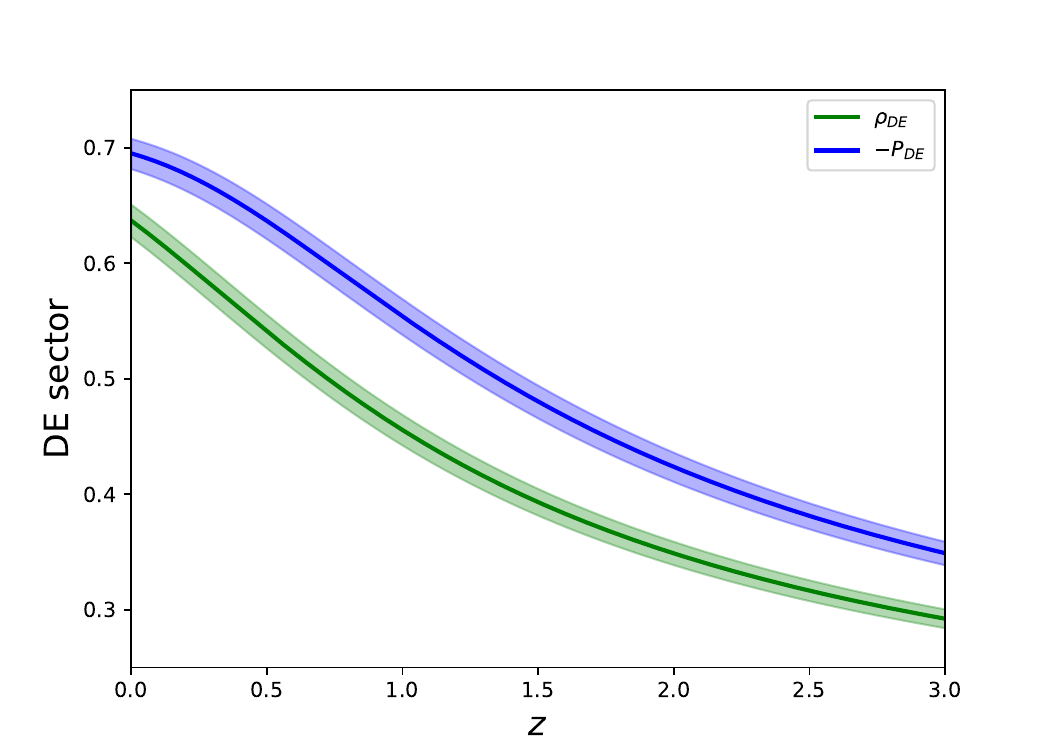}
	\caption{\label{figrpDE} The behavior of $\rho_{DE}$ and $-P_{DE}$ as functions of redshift $z$ for the LogDE model for the best fit values of the parameters as given by table \eqref{bestfit}. The shaded area denotes the $1\sigma$ error. }
\end{figure}

\begin{figure*}
	\includegraphics[scale=0.5]{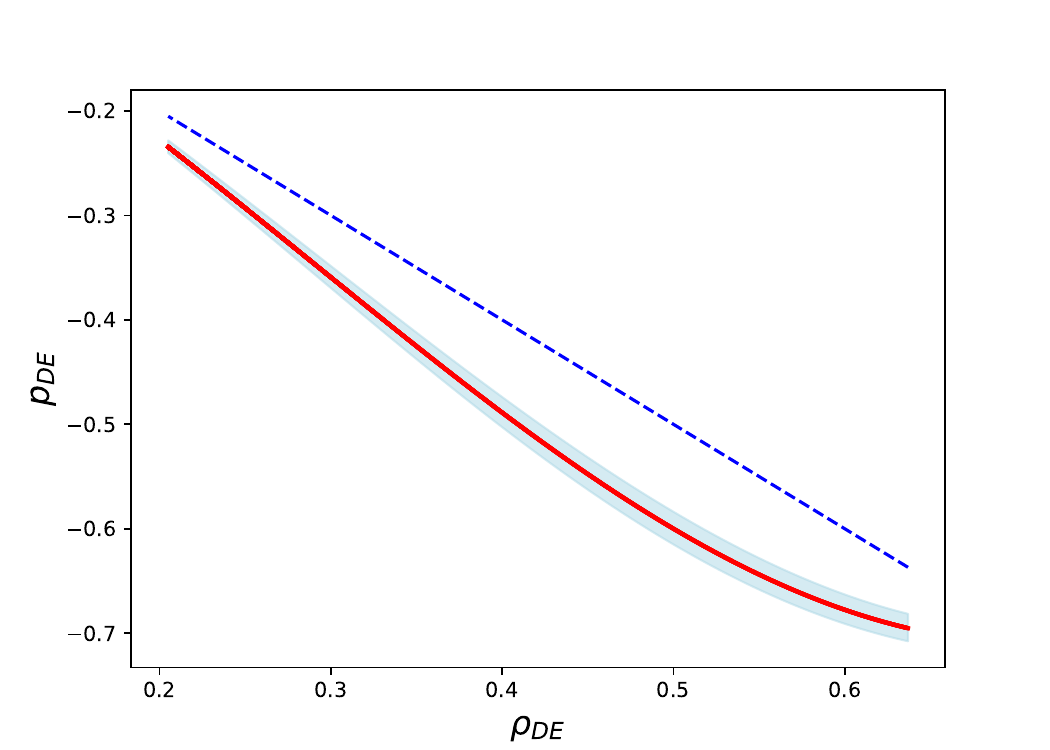}	\includegraphics[scale=0.455]{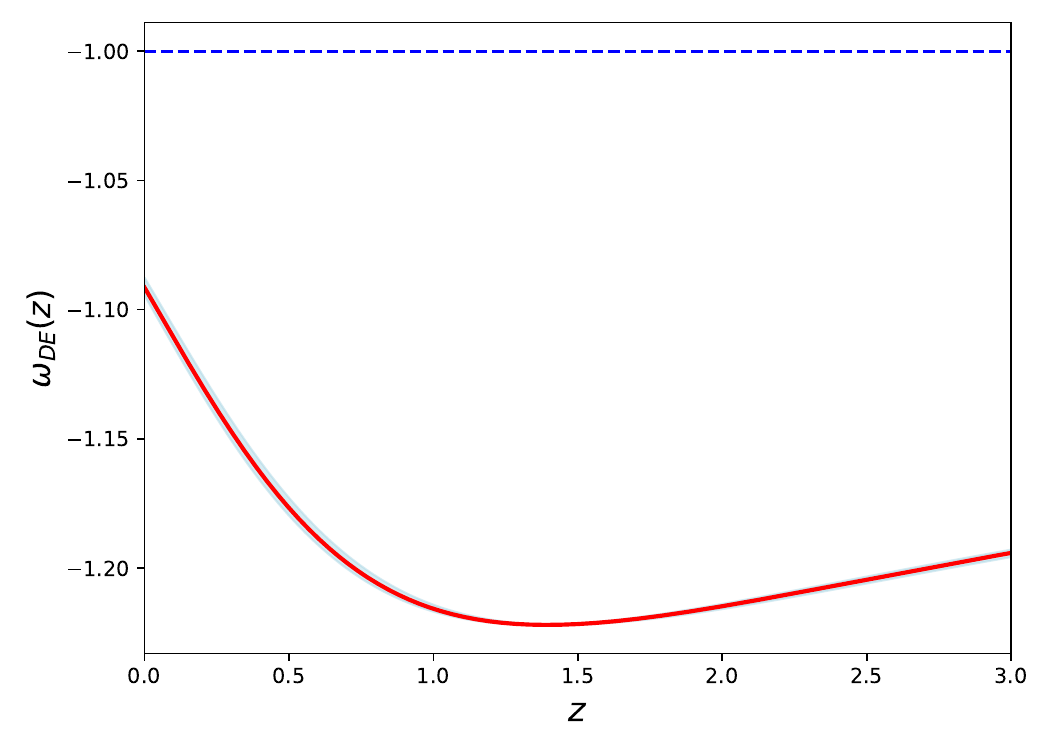}
	\caption{\label{figomegaDE} The behavior of $P_{DE}$ as a function of $\rho_{DE}$ (left) and the DE equation of state parameter $\omega_{DE}$ as a function of $z$ for the LogDE model for the best fit values of the parameters as given by table \eqref{bestfit}. The shaded area denotes the $1\sigma$ error. The dashed lines represent $\Lambda$CDM behavior.}
\end{figure*}

In order to find more about the dark energy sector, in figure \eqref{figrpDE} we have plotted the energy density of the DE sector together with the negative of the DE pressure. 
As one can see from the figure, the behavior of $\rho_{DE}$ and $-P_{DE}$ are very similar, signaling an approximate cosmological constant behavior for the LogDE. This is in fact in line with the previous plots, where the predictions of the logaarithmic DE model is very similar to that of $\Lambda$CDM. In figure \eqref{figomegaDE} we have plotted the behavior of the DE eos parameter, defined as
\begin{align}
	\omega_{DE} = \frac{\rho_{DE}}{P_{DE}},
\end{align}
together with the $\rho-P$ plane for both LogDE and $\Lambda$CDM models.
Both plots indicate that the DE sector behaves like phantom. However, the phantom line is very close to the cosmological constant. 

\begin{figure}[H]
	\includegraphics[scale=0.5]{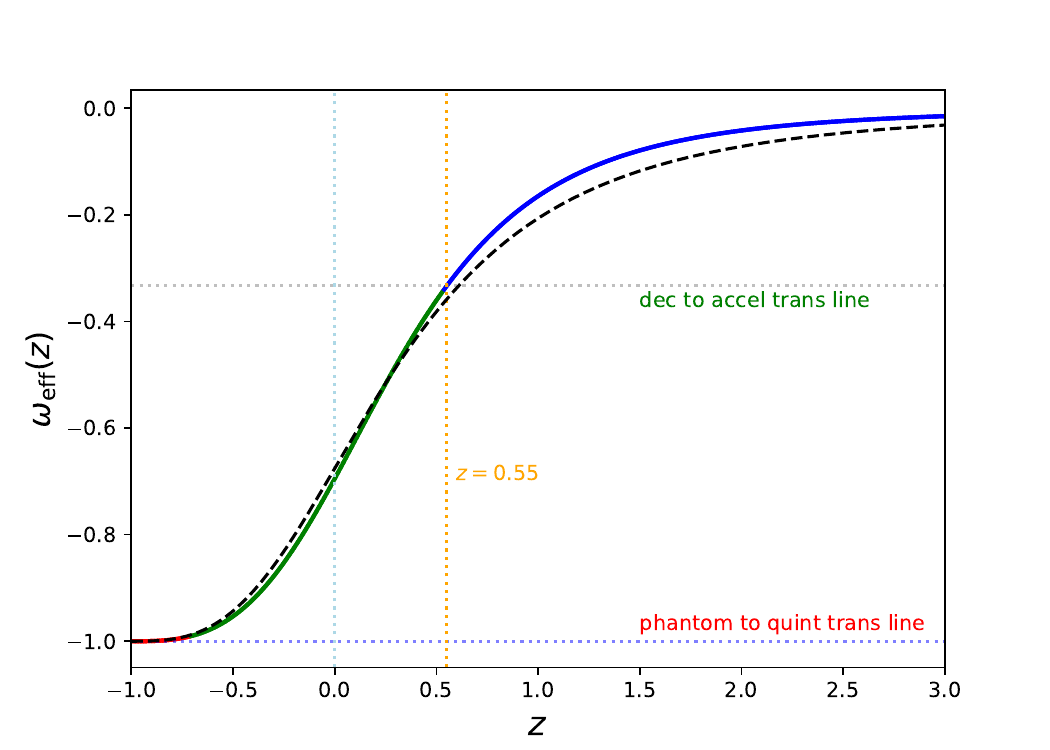}
	\caption{\label{figomegaeff} The behavior of $\omega_{eff}$ as a function of redshift $z$ for the LogDE model for the best fit values of the parameters as given by table \eqref{bestfit}. The shaded area denotes the $1\sigma$ error. Dashed line represents the behavior of the $\Lambda$CDM model.}
\end{figure}

The effective equation of state parameter
\begin{align}
	\omega_{eff}=\frac{P_{DE}}{\rho_m+\rho_{DE}},
\end{align} 
determines the total behavior of the universe. In figure \eqref{figomegaeff}, we have plotted the effective equation of state parameter $\omega_{eff}$ as a function of redshift $z$.
One can see from the plot that the total behavior of the universe is quintessence like, tending to the phantom behavior at far future. This is similar to the $\Lambda$CDM model except for the far future behavior which the $\Lambda$CDM model tends to the cosmological constant, never crossing the phantom line. As we have seen before, the other difference between the two models are the deceleration to acceleration transition time, which happens later for the LogDE model.
\section{Conclusions and final remarks}
In this paper we have explored the idea that the baryonic matter Lagrangian could be an arbitrary function of the energy density and pressure of the fluid. The energy-momentum tensor of a perfect fluid can be obtained from the matter Lagrangian of the form $\mathcal{L}_m=-\rho$ or equivalently $\mathcal{L}_m=P$. Despite the fact that these two choices result in the same energy-momentum tensor, these can lead to different theories when we have some non-standard matter coupling in the model like $f(R,T)$ gravities. So, it would be interesting to consider cases where the matter Lagrangian itself could be described using both cases.
We have seen that on top of FRW universe, in order to have a matter conservation on both baryonic and DE sectors, the function should be a linear combination of the pressure and an arbitrary function $B$ of the energy density $\rho$. 
This arbitrary function could account for the DE component. So, determining this function is equivalent to choosing a DE parameterization. The difference here is in the methodology. The parameterizations of DE usually assume dynamics for the eos parameter of DE through its dependence on the scale factor (redshift). Here, the parameterization of DE is done through its dependence on the energy density of the baryonic matter.

In order that the function $B$ correctly represents the DE, we have considered the energy conditions and also first order perturbations, to restrict $B$. We have found that $B$ should be positive, with negative first derivative and positive second derivative. Constant and power-law parameterizations are two important examples representing the $\Lambda$CDM and $\omega$CDM models, respectively.

Another example that meets above constraints is the logarithmic DE model, which we have explored in details in this paper. We have seen that this parameterization describes a phantom-like DE,  very close to the cosmological constant. We have then constrained the model parameters using two different combinations of the cosmic chronometers, Pantheom$^+$ and $f\sigma_{8}$ datasets. We have seen that the LogDE model predicts a slightly larger matter density abundance but otherwise identical to the $\Lambda$CDM model at least at late times. The early time behavior of the two models are different as the LogDE model predicts higher speed expansion, higher deceleration and lower growth rate. 

The DE component behaves very close to the cosmological constant but it belongs to the phantom regime. This implies that the behavior of the whole universe at far future becomes phantom-like which is in contrast to the standard behavior of the $\Lambda$CDM model.

In summary, besides this logarithmic DE example, the present framework could be used to parameterize the DE component through its dependence on the baryonic energy density of the universe, providing a unified framework for both baryonic and DE sectors.

\end{document}